\documentstyle[12pt,aasms4]{article}

\received{}
\accepted{}

\begin{document}
\title { Ultra-High-Energy Cosmic Ray Acceleration by Magnetic 
Reconnection in Newborn Accretion Induced Collapse Pulsars}

\author{Elisabete M. de Gouveia Dal Pino\altaffilmark{1,2}
\& Alex Lazarian\altaffilmark{3} } 
\altaffiltext{1}{Instituto Astron\^omico e Geof\'{\i}sico, University 
of S\~ao Paulo, Av. Miguel St\'efano, 4200, S\~ao Paulo
04301-904, SP, Brasil; 
E-mail: dalpino@iagusp.usp.br,  } 
\altaffiltext{2}{
Astronomy Department,
Theoretical Astrophysics Center,
University of California,
601 Campbell Hall,
Berkeley, CA 94720-3411} 
\altaffiltext{3}{Department of  Astronomy, 
University of Wisconsin, 
Madison, USA; 
E-mail: lazarian@dante.astro.wisc.edu }

\begin{abstract}
We here investigate the possibility that the 
ultra-high energy  cosmic ray (UHECR) events observed above the GZK 
limit are mostly protons accelerated in reconnection sites just above 
the 
magnetosphere of newborn millisecond pulsars 
which are originated by accretion 
induced collapse (AIC). 
We formulate the requirements for the acceleration mechanism
and show that
AIC-pulsars with surface
magnetic fields 
 $10^{12} $ G  $<  \, B_{\star} \lesssim  $ $10^{15}$ G 
and spin periods
1 ms $\lesssim \, P_{\star} \, < \, $ 60  ms, 
are able to accelerate
particles to energies $\geq \, 10^{20} $ eV.
%
Because the expected rate of AIC sources in our Galaxy 
is very small ($\sim \,  10^{-5}$ yr$^{-1}$),  
the corresponding contribution to
the 
flux of UHECRs is  neglegible, 
and the total flux 
is given  by the integrated contribution from  AIC sources 
produced by the distribution of galaxies located within the distance 
which 
is unaffected by the GZK cutoff ($\sim \, 50 $ Mpc). 
We find that reconnection 
should convert a fraction 
%
$ \xi \, \gtrsim \, 0.1$ 
of magnetic energy
into UHECR
in order to reproduce the observed flux.

\end{abstract}


\section{Introduction}
The detection of  cosmic ray events with energies beyond 10$^{20}$eV
by AGASA (Takeda et al. 1999), Fly's Eye (Bird et al. 1995), and 
Haverah 
Park
(Lawrence, Reid, \& Watson 1991)
experiments still poses a challenge for the understanding of their 
nature
and sources. These ultra-high energy cosmic rays (UHECRs) show no major
differences in their air shower characteristics to cosmic rays at lower
energies and thus one would expect them to be mostly protons 
(Protheroe
1999). 
If UHECRs are charged particles, or 
protons, then they should be 
affected 
by
the expected Greisen-Zatsepin-Kuzmin (GZK) energy cutoff ($\sim 5\times
10^{19}$ eV), which is due to photomeson production by interactions 
with 
the
cosmic microwave background radiation, unless they are originated at
distances closer than about 50 Mpc  (e.g., Protheroe \& Johnson 1995,
Medina Tanco, de Gouveia Dal Pino \& Horvath 1997). On the other hand, 
if
the UHECRs are mostly protons from nearby sources 
(located within $\sim $
 50 Mpc), then the arrival directions of the events should point toward
their sources since they are expected to be little deflected by the
intergalactic and Galactic magnetic fields (e.g., Stanev 1997, Medina 
Tanco,
de Gouveia Dal Pino \& Horvath 1998). The present data shows no 
significant
large-scale anisotropy in the distribution related to the Galactic disk 
or
the local distribution of galaxies, although some clusters of events 
seem 
to
point to the supergalactic plane (Takeda et al. 1999, 
Medina Tanco 1998). 

A number of source candidates and acceleration mechanisms have been 
invoked but all of them have their shortcomings 
(see, e.g.,
%
Olinto 2000 for a review).
%
%
%
Particles can,
in principle, extract the required energies from an induced e.m.f. in a
circuit connected between the polar and the last open field line of a 
rapidly
rotating pulsar, although it is not clear how the large voltages  can 
be maintained (e.g., Hillas 1984), or be accelerated in reconnection
sites of magnetic loops, $if$ these can be produced, e.g., 
by Parker instability, 
on the surface of a pulsar (Medina Tanco, de Gouveia Dal Pino \& 
Horvath 
1997), but the accelerated particles will probably lose
most of their energy gain by curvature radiation while dragged along by 
the magnetic dipole  field (Sorrell 1987).  Alternatively, Olinto, 
Epstein, \& Blasi (1999) have recently proposed that UHECRs could be 
iron nuclei
stripped by strong electric fields from the surface of highly 
magnetized
neutron stars and accelerated in a relativistic MHD wind.  It is 
not clear however,  how the accelerated particles can
escape from the magnetosphere of the star, for although efficient power
 extraction may be
possible, there is a dense positron-electron plasma generated with the
relativistic wind (Gallant \& Arons 1994) 
that will possibly modify the electric fields and also
interact with the energetic particles.  Their model also predicts that 
a
correlation with the Galactic plane should become evident as data 
collection
at the highest energies improves.
In this paper we discuss an alternative model in which UHECRs are 
accelerated in
magnetic reconnection sites outside the magnetosphere of  very young 
millisecond pulsars being produced by accretion induced collapse (AIC) 
of a 
white dwarf.
\section{The model}
When a white dwarf reaches the critical Chandrasekhar mass $\sim 1.4$
M$_{\odot }$ through mass accretion, in some cases it does not explode 
into
a type Ia supernova, but instead collapses directly to a neutron star 
(e.g.,
 Woosley \& Baron 1992 and references therein). 
The accretion flow spins up the star and confines the magnetosphere  
to a radius $R_X$ where  plasma stress in the accretion disk  and 
magnetic 
stress 
balance
(Arons 1993). At this radius, which also defines the inner radius of 
the 
accretion disk,   the equatorial flow will divert into a funnel inflow
along the closed 
field-lines toward the star (Gosh and Lamb 1978), and 
a centrifugally 
driven wind outflow  (Arons 1986). 
Recently, Shu et al. (1994, 1999)
have studied the detailed field geometry 
of  magnetized stars accreting matter from a disk.  
Two surfaces of null poloidal field lines are required
to mediate the geometry of dipole-like field lines of the star with 
those
opened by the wind and those trapped by the funnel inflow emanating 
from
the $R_X$ region. Labeled as "$helmet$ $streamer$" and "$reconnection$ 
$ring$" 
in
Figure 1, these magnetic null surfaces begin or end on $Y$ points. 
Across each  null surface, the poloidal field suffers a sharp reversal
of direction. 
Dissipation of 
the large 
electric currents that develop along the null surfaces
will lead to reconnection of the oppositely  directed 
field lines (e.g., Biskamp 1997, Lazarian \& Vishniac 1999).
Helmet streamers (or flare loops) are also present in the magnetic 
field configuration
 of the 
solar corona. 
The magnetic
energy released by reconnection in the helmet streamer 
 drives violent
outward motions in the surrounding plasma
that accelerate copious 
amounts 
of solar cosmic rays without producing many photons (Reames 1995).
A similar process may take place in the helmet streamer of  
young born AIC-pulsars and the magnetic energy released may 
accelerate particles to the UHEs.
The  reconnection rings in the disk  would
also, in principle,  be able to accelerate UHECRs, 
however, as we  will see below,
the intense radiation field  produced in the disk must 
prevent  accelerated UHE particles to escape from it.
The particular mechanism of particle acceleration
during the reconnection events is still unclear in spite of 
numerous attempts
to solve the problem 
(see 
LaRosa et al 1996,  Litvinenko 1996).
Cosmic rays from the Sun confirm that the process is sufficiently 
efficient in spite of the apparent theoretical difficulties for its
explanation.
Solar flares observations  indicate that 
the reconnection speed 
can 
be 
as high as one tenth of the Alfv\'en velocity. 
As in the conditions we deal with this speed
approaches $c$, the expected acceleration rate is large. 
We discuss the details of  particle acceleration  during 
reconnection events in  
Lazarian \&  de Gouveia Dal Pino (2000). 
We argue  that 
 protons can be accelerated by the large induced electric field
within the reconnection region  
(e.g., Haswell, Tajima, 
\& Sakai 1992, Litvinenko 1996) over a time scale 
$\sim \Delta R_X/c$,
where $\Delta R_X$ is the size of the reconnection zone (see
below).
The required electric field is less than the critical value for pair
production and therefore is sustainable (see also \S 3).
%

For a Keplerian disk, the inner disk edge $R_X$ rotates at 
an angular speed
($G M_{\star}/R_X^3)^{1/2}$, and  equilibrium between gravity 
and centrifugal
force at $R_X$ will lead to co-rotation of the star
with the inner disk edge, i.e., 
$R_X \, = \, (G M_{\star}/\Omega _{\star}^2)^{1/3}$,
which for typical millisecond pulsars with 
rotation periods 
$P_{\star} = 2\pi/\Omega_{\star} \simeq 1.5 - 10$ ms,  mass
$M_{\star} \, \simeq  \, $1 M$_{\odot}$, 
and  radius $R_{\star} = 10^6$ cm, 
gives 
$R_X \simeq (2 $ to $ 7) \times 10^{6}$ cm $R_{6}$,
where $R_{6} = R_{\star}/ 10^6$ cm.

The primary condition usually assumed 
on the region 
to accelerate particles of charge $Ze$ to energies $E$ 
 is that its  width
$\Delta R_X \,  \geq \, 2 \, r_L$, 
where 
$r_L$ is the particle Larmour radius 
$r_L = E /Z e \, B_X$ (Hillas 1984)  and
$B_X$  is the magnetic 
field (normal to particle velocity) at the $R_X$ region,
$B_X \, \simeq \, B_{dipole}(R_X) \left( {\frac{R_X}{ \Delta R_X} 
}\right)^{1/2}$
(e.g., Arons 
1993), where  $B_{dipole}(R_X) = B_{\star} \, (R_{\star}/R_X)^{3}$ is the 
magnetic field that would be
present in the absence of the shielding disk, and 
$B_{\star}$ is the magnetic field at the surface 
of the star.  
While this condition on $\Delta R_X$ is usually invoked to allow
 particles to bounce back and forth thus gaining energy, we find
that for accelerated protons the synchrotron losses
may be too large if they bounce within the reconnection
zone. Then, in our model  
the condition above is invoked to assure that the 
field $B_X$ will focus
particles to move within a small angle into the reconnection zone.
%
Besides, 
one 
should also expect that: 
$\Delta R_X/ R_X << 1$. Both conditions above  imply that
$1 \, > > \, \left({\frac{\Delta R_X}{R_X } }\right) \, \geq \, 4 e^{-
2} 
\, Z^{-2} \, R_{\star}^{-6} \, (G \, M_{\star})^{4} \,  
E^{2} \, B_{\star}^{-2} \, \Omega_{\star}^{-8/3}$. 
This relation indicates that for a given ratio $\Delta R_X/R_X$ and 
particle 
energy $E$, the stellar magnetic field $B_{\star}$ must 
satisfy 
\begin{equation}
B_{13} \, \gtrsim \, Z^{-1} \,  E_{20} \, \Omega_{2.5k}^{-4/3} \, 
   \left({\frac{\Delta R_X/R_X } { 0.1}}\right)^{-1/2} 
\end{equation}
\noindent 
where we have assumed 
$M_{\star} \, =  \, 1  M_{\odot}$, 
$R_{\star} = R_6$,
$E_{20} \, = \, E/10^{20} $ eV, 
$\Omega_{2.5k} = \Omega_{\star}/ 2.5 \times 10^3$ s$^{-1}$,
and $B_{13} = B_{\star}/10^{13}$ G.
The corresponding allowed zones in the $B_{\star} - \Omega_{\star}$
plane are shown in Figure 2 for $E_{20} = 1$ and $E_{20} = 10$,
and different values of the ratio $\Delta R_X/R_X$. 
The curves with $\Delta R_X/R_X = 1  $ determine extreme lower bounds 
on the stellar surface magnetic field and the angular speed.  
We note that  stellar magnetic fields 
 $10^{12} $ G $ <  B_{\star} \lesssim  $ $10^{15}$ G 
and angular speeds 
$4 \times 10^{3}$ s$^{-1}$ $\gtrsim \Omega_{\star} \, > \,  10^{2} $ 
s$^{-1}$,
are able to accelerate
particles to energies $E_{20} \,  \gtrsim $ 1. 
The values above are perfectly compatible with the parameters 
of young pulsars and
Eq. (1) is thus a good representation of the typical conditions 
required 
for particle acceleration to the UHEs in reconnection zones of AIC-
pulsars.
%
%
The substitution of Eq. (1) into the equation for $B_X$ 
implies a magnetic field 
in  the acceleration zone 
$B_X \, \simeq \,  1.5 \times 10^{12}$ G $B_{13} \, \Omega_{2.5k}^2 \, 
\left({\frac{\Delta R_X/R_X } { 0.1}}\right)^{-1/2}$.

A newborn millisecond pulsar spins down due to 
magnetic 
dipole radiation in a time scale given by
$\tau_{\star} = \Omega_{\star}/\dot \Omega_{\star} \simeq 
\left({\frac{ I c^3 }{B_{\star}^2 R_{\star}^6 \Omega_{\star}^2 } 
}\right)$, 
which for a moment of inertia 
$I = 10^{45}$ g cm$^2$
gives
$\tau_{\star} \simeq 4.3 \times 10^7 $ s $  B_{13}^{-2} \, 
\Omega_{2.5k}^{-2}$. 
We can show that the 
condition that the magnetosphere and  
the disk stresses are in equilibrium 
at 
the inner disk edge results a disk mass accretion rate 
$\dot M_D \, \simeq \,  \alpha_X^{-2} \, I c^3  \, 
(G \, M_{\star})^{-5/3} \, \Omega_{\star}^{1/3} \, 
\tau_{\star}^{-1} $, 
where 
$2 \gtrsim \alpha_X > 1$
 measures the amount of 
magnetic dipole flux that has been pushed by the disk accretion flow to 
the inner edge of the disk 
(Gosh \& Lamb 1978, 
 Shu et al. 1994). 
Substitution of the 
previous equations yields 
$\dot M_D \, \simeq \,  3 \times 10^{-8} M_{\odot} $ s$^{-1} \, 
{\alpha_2}^{-2} \,
B_{13}^2 \, \Omega_{2.5k}^{7/3}$ 
(where $\alpha_2 = \alpha_X/2$),
  which is much larger than the 
Eddington accretion rate 
$\dot M_{Edd} \simeq 7.0 \times 10^{-17} M_{\odot} $ s$^{-1}
(M_{\star}/ M_{\odot})  $.
However, this "super-Eddington" accretion 
(which is correlated to $\tau_{\star}$) will 
last for a time  $\tau_{D}$, which is only
a small fraction ($f_D$) of  $\tau_{\star}$. 
The strong radiation pressure from the accreted material will  
 cause most of the infalling material to be ejected from the system,
and this will in turn cause the accretion rate 
to rapidly decrease to
a value nearly equal to the Eddington limit
(e.g., Lipunova
1999). 
Advection dominated inflow-outflow solutions involving 
supercritical accretion onto neutron stars predict a 
total mass 
depostion on  the star 
$\sim $  few $0.01 M_{\odot}$ (e.g., Brown et al. 1999). 
Thus assuming that a mass $M \sim 0.04 M_{\odot}$ 
is accreted during the supercritical phase, 
we find 
$\tau_D \simeq M/\dot M_D \simeq 1.3 \times 10^6 $ s, and
$f_D \simeq 0.03$.     
The acceleration of UHECRs in the reconnection zone, 
on the other hand, will 
last as long as the
supercritical accretion. 
It is  well known that the most violent solar flares 
can live up to several hours and the more energetic the 
longer-lived they are. In our scenario, 
considering that the reconnecting magnetic fields 
are several orders of magnitude larger than
in the solar corona, 
it is reasonable to expect that
the most violent reconnection events can live
at least for several days, 
as required by $\tau_D$. 
The spectrum evolution of 
the 
accelerated UHECRs will be, therefore, determined by 
$\tau_D = f_D \tau_{\star}$ 
(see below). 

In order to derive the spectrum of accelerated particles, let us first 
evaluate the rate of magnetic energy that can be extracted from the 
reconnection region, 
$\dot W_B \simeq (B_X^2/ 8\pi) \, \xi v_A \, (4\pi R_X \,\Delta R_X 
)$,
where $v_A \sim \, c$ is the Alfv\'en velocity,
and $\xi \, < \, $1 is a factor that determines the amount of magnetic 
energy released in the reconnection that will  accelerate the 
particles. 
Substituting the previous relations into this equation, one finds
$\dot W_B \, \simeq \,  2.6 \times 10^{46} $  erg  s$^{-1} \, \xi 
\, B_{13}^2 \, \Omega_{2.5k}^{8/3}$. 
According to our model assumptions, in a  reconnection burst an electric
field arises inductively 
because of plasma fow across the $\vec B$ lines. The corresponding
maximum voltage drop is
$ V \, \simeq \, \epsilon \, \Delta R_X \, \simeq (v_A/c) B_X  \Delta R_X 
\simeq B_X  \Delta R_X  $
where $\epsilon$ is the electric field strength 
(e.g., Bruhwiller and Zweibel 1992).
Once a particle decouples from the injected fluid, it 
will be ballistically accelerated by the electric field and the
momentum attained  will depend on the intensity of the electromagnetic
burst and on the time and location at which it decouples from the
injected fluid motion (e.g., Haswell et al. 1992).
It is out of the scope of this work to quantitatively constrain the
burst characteristics and thus to deduce the detailed spectrum of particle
energies. Clearly, however, the more energetic bursts 
will give rise to a 
higher  average energy for the ejected particles 
(Haswell et al. 1992). In these extreme cases, the induced electric 
voltage will dominate and the particles will be accelerated to an 
average energy 
$E \simeq  V \,  Z e $, which 
according to the equation above and consistently with 
relation (1) is 
$E \simeq 10^{20}$ eV.
The UHECR flux emerging from the reconnection site can then be 
estimated as
$\dot N \, \simeq \, {\frac {\dot W_B } { E}}  \, \simeq \, 1.6 
\times 10^{38} \, {\rm s}^{-1} \,  \xi \, B_{13}^2 \, 
\Omega_{2.5k}^{8/3} \,E_{20}^{-1}$
for particles with energy $E \gtrsim  10^{20} $ eV, and
 the particle spectrum $N(E)$ is obtained from 
$\dot N \, = \, N(E) \, {\frac {dE } { dt}}  \, 
\simeq \, N(E) \,  {\frac {dE } { d \Omega_{\star} }  } \, \dot 
\Omega_{\star}/f_D$,
or 
\begin{equation}
N(E) \, \simeq 
\, 1.7 \times 10^{33} \, {\rm GeV}^{-1} \,  \xi \, Z^{-1/2} \,  
B_{13}^{-
1/2} \, E_{20}^{-3/2} \, \left({\frac{\Delta R_X/R_X } 
{0.1}}\right)^{-1/4}
\end{equation}
\noindent 
where Eq. (1) 
 has 
given 
$ d \Omega_{\star}/dE  \simeq 2.1 \times 10^{-17} \, Z^{-3/4} \, 
E_{20}^{-1/4} \, B_{13}^{-3/4} \,  \left({\frac{\Delta R_X/R_X } { 
0.1}}\right)^{-3/8}$ 
 (with the signal made equal in Eq. 1). 
Eq. (2) above predicts that 
$N(E) \propto E^{-3/2} = E^{-1.5}$, which is a flat spectrum
%
in 
good
agreement with observations (e.g., Olinto 2000).

The particle distribution emerging from the source will not be 
isotropic. Thanks to the magnetic field geometry in the reconnection 
site (see Fig. 1), it will be confined to a  ring (above and below the 
accretion disk)
of thickness and 
height  given both by $\sim \Delta R_X$.  
The distribution will thus be beamed in  a solid angle 
$\Delta \Omega \, \simeq \, 4 \pi (\Delta R_X/R_X)^2$.
Let us now, estimate the resulting flux of UHECRS at the Earth. 
The total number of objects formed via AICs 
in our Galaxy is limited by nucleosynthesis constraints to a very small 
rate  
$ \sim \, (10^{-7} \, - \,  10^{-4}$) yr$^{-1}$, 
or in other words, less 
than 0.1 \% of 
the total 
galactic neutron star population (Fryer et al. 1999).  
Assuming then that the  rate of AICs in the Galaxy is
${\tau_{AIC} }^{-1} \, \simeq  \, 10^{-5}$ yr$^{-1}$, we can evaluate 
the probability $a $ $priori$ of having UHECRs events produced in the 
Galaxy. 
The beaming  mentioned above
will reduce the probability
of detection of the events of a source by a factor 
$f_b \sim \,  (\Delta R_X/R_X)^2 \simeq 10^{-2}$.
Thus the probability 
will be only
$P \, \simeq \, f_b \, {\tau_{AIC}}^{-1}  \, t  \, \simeq \, 2 \times  
10^{-
6}$, where $t = 20 $
years accounts for the time the UHECR events have been collected  in 
Earth detectors since the operation  of the first experiments. 
Since the individual contribution to the observed UHECRs due to  
AICs in our  Galaxy is so small we must evaluate the integrated 
contribution due to  AICs from all the galaxies located within a
volume 
which is not affected by the GZK effect, i.e., within a radius 
$R_{50} = R_G/ 50$ Mpc.
Assuming  that each galaxy has essentially the
same rate of AICs as our Galaxy and taking the standard galaxy 
distribution 
$n_G \simeq \, 0.01 \, e^{\pm 0.4}\, h^3 $ Mpc$^{-3}$ (Peebles 1993)
(with the Hubble parameter defined as $H_o = h$ 100 km s$^{-1} $ 
Mpc$^{-1}$), the resulting flux at $E_{20} \, \geq $  1 is
$F(E) \, \simeq \, \,  N(E) \, n_G \, {\tau_{AIC}}^{-1} \, R_{G}$, 
which gives
\begin{equation}
F(E) \, \simeq \, 3.3 \times 10^{-29} \xi \, {\rm GeV}^{-1} 
{\rm cm}^{-2} {\rm s}^{-1} \,
Z^{-1/2} \, B_{13}^{-1/2} \, 
E_{20}^{-3/2} \, {\tau_{AIC,5}}^{-1} \, n_{0.01}  \, 
R_{50} \left({\frac{\Delta R_X/R_X } {0.1}}\right)^{-1/4} 
\end{equation}
\noindent 
where ${\tau_{AIC,5}}^{-1} \, = \, {\tau_{AIC}}^{-1}/ 10^{-5}$ yr$^{-
1}$,
 and  
$n_{0.01} = n_G/0.01 $ h$^3$ Mpc$^{-3}$.
Observed data by the AGASA experiment (Takeda et al. 1999) gives a flux 
at 
$E = $10$^{20}$  eV  of 
$F(E) \, \simeq  \,  4 \times \, 10^{-30}$ Gev$^{-1}$ cm$^{-2}$ s$^{-
1}$, so 
that the efficiency of converting magnetic energy into UHECR should be 
$ \xi \, \gtrsim \, 0.1$
in order to reproduce such a signal.
%
%
\section{Conclusions and Discussion}
We have discussed the possibility that the UHECR events observed above 
the GZK limit are protons accelerated in reconnection sites just above 
the magnetosphere of very young  millisecond pulsars originated by 
accretion induced collapse.
AIC-pulsars with surface
magnetic fields 
 $10^{12} $ G  $<  \, B_{\star} \lesssim  $ $10^{15}$ G 
and spin periods
1 ms $\lesssim \, P_{\star} \, < \, $ 60  ms, 
are able to accelerate
particles to energies $\geq \, 10^{20} $ eV. These limits can be 
summarized by the condition 
$B_{\star} \,  \gtrsim \, 10^{13}$ G $  (P_{\star}/2.5 $ ms)$^{4/3}$  
(Eq. 1 and 
Fig. 2).
Because the expected rate of AIC sources in our Galaxy 
is very small, 
the total flux is 
given  by the integrated contribution from  AIC sources produced in the 
distribution of galaxies within a volume which is unaffected by  the 
GZK cutoff (of
radius $R_G \simeq 50 $ Mpc). 
We find that the reconnection efficiency factor needs to be 
$ \xi \, \gtrsim \, 0.1$
in order to reproduce the observed flux.
This result is appealing because it predicts 
no correlation of UHECR events with the Galactic plane, in agreement 
with present observations. 
%
However,  as data collection improves, we 
should expect some sign of correlation with the local distribution of 
galaxies and the supergalactic plane. 
These predictions 
can be ultimately tested by coming experiments such as 
the AUGER Observatory which will provide high statistics samples of 
UHECRs.


The model 
predicts a highly super-Eddington accretion 
mass 
rate  during part of the AIC process. In such a regime, one 
should 
expect a large mass outflow from the stellar surface itself
which might 
alter the coronal conditions near the helmet streamer region. However,
the strong closed magnetic fields in the pulsar magnetosphere can 
inhibit 
the gas 
outflow from the stellar surface if the 
surface temperature does not exceed the value at which the radiation 
energy 
density  ($a \, T^4$) equals the magnetic field energy density 
($B_{\star}^2/8 \pi$), i.e., 
$T_{\star} \lesssim 5 \times 10^9 $ K $ B_{13}^{1/2}$; this upper limit 
is 
consistent with the predicted values for a neutron star that is formed
by AIC (Woosley \& Baron 1992).

Let us consider now the energy loss mechanisms that may 
affect the efficiency of the acceleration of the particles to the UHEs. 
Energy losses by curvature radiation which  occur when cosmic rays have to 
stream along magnetic field lines with a finite radius of curvature 
$R_c$, have 
a  cooling time 
$t_{rc} \propto  R_c^2 $. 
In our model, the reconnection takes place in the helmet streamer 
region
(see Fig. 1) where  $R_c \rightarrow \infty$ and 
therefore, $t_{rc} > > t_a$, where 
$t_a$ is the acceleration time for a proton in the reconnection site.
[We can 
use the dimensions  of the reconnection site derived in \S 2 
to estimate the order of magnitude of 
$t_a \simeq \Delta R_X/ c \, \simeq 8 \times 10^{-6} {\rm s} \,
\left({\frac{\Delta R_X/R_X}{0.1} }\right) \, \left({ \frac{R_X} {2.45 
\times 10^6  {\rm cm} } }\right)$.]
The time for synchrotron losses  is  
$t_{syn} \propto \theta^{-2}$, where  
 $\theta$ is the particle pitch angle. 
 Since we require that the protons in the reconnection zone
 are accelerated by an induced electric field 
rather than by a scattering process, 
they are expected to 
maintain 
beamlike pitch angles while escaping along the magnetic field lines,
 i.e., 
$\theta < < 1$  and $t_{syn} > > 
t_a$. 
%

The accelerated protons may also undergo  energy loss by pion 
and $e^{\pm}$ 
pair production 
due to interactions with photons from the 
accretion disk radiation field.
The characteristic distance scales for these processes 
can be estimated by  
$\lambda_{p \gamma} \simeq (\sigma_{p \gamma} n_{\gamma})^{-1}$  and 
$\lambda_{pair} \simeq (\sigma_{pair} n_{\gamma})^{-1}$ 
for 
photo-pion  and pair production, respectively,
where $n_{\gamma}$ is the photon number 
density at the reconnection site, 
$\sigma_{p \gamma} \simeq  2.5 \times 10^{-28} $ cm$^{-2}$ is the cross 
section 
for pion production, 
 and 
$\sigma_{pair} \simeq  10^{-26} $ cm$^{-2}$ is the cross section 
for pair production 
(e.g., 
Bednarek and Protheroe 1999). 
In order to estimate $n_{\gamma}$,
we must determine the luminosity in the accretion disk. 
At super-Eddington accretion rates, the disk is thicker 
and heat is radially advected 
with matter.  The luminosity of such advective supercritical accreting 
disks
is given by 
$L_D \simeq [0.6 + 0.7 \, ln (\dot M_D/\dot M_{Edd})] \, L_{Edd}$
(e.g., Lipunova 1999),
where 
$L_{Edd} \simeq 1.25 \times 10^{38} $ erg s$^{-1}$ $(M_{\star}/ 
M_{\odot})$.
Using the value obtained in \S 2 for 
$\dot M_D/ \dot M_{Edd} \simeq 1.8 \times 10^{9}$, we
find $L_D \simeq 1.8 \times 10^{39}$ erg s$^{-1}$.
Inside the disk, where the optical depth is much larger than unity, 
photons 
and particles are in thermodynamic equilibrium and the disk radiates 
like a 
black-body with a temperature 
$T_D \simeq 1.1 \times 10^9 $ K  
$\left({ \frac {\dot M_D } {3 \times 10^{-8} M_{\odot} \, {\rm s}^{-
1} } }\right)^{1/4} \,
\left({ \frac{ M_{\star}} {M_{\odot} } }\right)^{1/4} \, 
\left({ \frac{ R_X} {2.45 \times 10^6  {\rm cm} } }\right)^{-3/4}$.
At a distance $R_X$, 
$n_{\gamma} \simeq L_D/ (\pi R_X^2 \bar \epsilon_{\gamma})$
(e.g., Sorrell 1987),
where $\bar \epsilon_{\gamma} \, \simeq \, 2.8 k_B T_D \sim 0.35 MeV$ 
is the 
mean photon energy, or
 $n_{\gamma} \simeq 8 \times 10^{21} {\rm cm}^{-3} \,
\left({ \frac{ L_D} { 1.9 \times 10^{39} {\rm erg} \, {\rm s}^{-1} } }\right) \,
\left({ \frac{ R_X} {2.45 \times 10^{6}  {\rm cm} } }\right)^{-2} \, 
\left({ \frac{ T_D} { 1.5 \times 10^9  {\rm K} } }\right)^{-1}$. 
We note that beyond the disk, 
the luminosity can be substantially  smaller as the matter 
just above the disk surface  can be partially opaque to the outgoing 
radiation. 
Thus, the value above
of $n_{\gamma} $ for the
helmet streamer is probably overestimated. 
Substitution of $n_{\gamma}$ 
into the equation for $\lambda_{p \gamma}$ gives
$\lambda_{p \gamma}/R_X  \, \gtrsim  \, 0.22 \, > 0.1 
\left({\frac{\Delta R_X/R_X } { 
0.1}}\right)$, 
so that the 
interaction distance scale for pion production 
is 
larger than the width of the acceleration region. For pair production, 
although the interaction distance $\lambda_{pair} $ is shorter, 
the inelasticity of this process is so small 
(inelasticity $K_p \sim \, 2 \times 10^{-3}$) that the corresponding 
energy 
loss rates of the UHECR protons are even smaller than in the case of 
pion production (e.g., Bednarek \& Protheroe 1999).
However, the injected $e^{\pm}$ pairs can initiate inverse Compton pair 
cascades in the reconnection region that may saturate an induced 
electric field if the width of the reconnection region is larger than 
$\lambda_{sat} \simeq 100 \, (\sigma_{trip} n_{\gamma})^{-1}$,
where $\sigma_{trip} \simeq  1.3 \times 10^{-26} $ cm$^{-2}$ 
is the cross section 
for the triple $e^{\pm}$ pair production by electron-photon collisions. 
We find that
$\lambda_{sat}/R_X \gtrsim 0.4 > 0.1 \left({\frac{\Delta R_X/R_X } { 
0.1}}\right)$, 
so that the width of the reconnection site is smaller than 
$\lambda_{sat}$.


\acknowledgements
This paper has benefited from many valuable comments 
of 
an 
anonymous referee, and 
J. Arons, A.E. Glassgold, J. A. de Freitas Pacheco,
J. Horvath, A. Melatos, and F.H. Shu.   
E.M.G.D.P. has been partially supported by a grant of the Brazilian 
Agency FAPESP.

\newpage
{\bf Figure Caption}

\noindent
Figure 1. Schematic drawing of the magnetic field geometry and the gas 
accretion flow in the inner disk edge at $R_X$. UHECRs are accelerated 
in 
the magnetic reconnection site at the helmet streamer (see text). The 
figure also indicates that coronal winds from the star and the disk 
help the magnetocentrifugally driven wind at $R_X$ to open the field 
lines 
around the helmet streamer (adapted from Shu et al. 1999).  

\noindent
Figure 2. Allowed  zones in the parameter space for UHECR
acceleration for 
$E = 10^{20}$  and $10^{21}$ eV. The vertical line in each plot
indicates the upper 
limit on
the stellar angular speed 
$\Omega_{\star, max}  \, = \,  (G \, M_{\star}/R_{\star}^3)^{1/2} \, 
\simeq 1.15 \times 10^4 $ s$^{-1}$. The allowed zone in each plot 
is above the solid line
for which 
$\Delta R_X/R_X = 1$; the dash-dotted line has 
$\Delta R_X/R_X = 0.1$; and the dashed line has
$\Delta R_X/R_X = 0.01$.

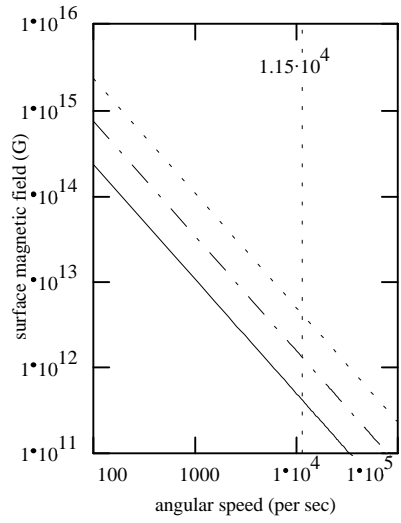
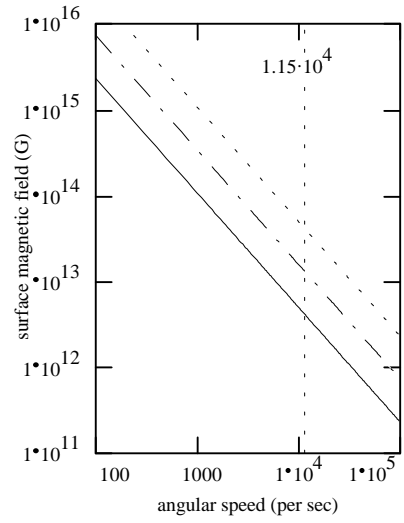
\end{document}